\documentclass[a4paper,11pt,twoside]{article}

\usepackage{geometry} 
\usepackage{amsmath}
\usepackage{amssymb}
\usepackage{color}
\usepackage{graphicx}
\usepackage{verbatim}

\newcommand{\TeV}{\,\mathrm{TeV}}
\newcommand{\GeV}{\,\mathrm{GeV}}

\newcommand{\MGUT}{M_\text{GUT}}

\newcommand{\ord}[1]{\mathcal{O}\left( #1 \right)}

\newcommand{\eq}[1]{eq.~(\ref{eq:#1})}

\newcommand{\omi}[1]{}

\newlength{\myem}
\newcounter{mysubequation}[equation]

\makeatletter
\renewcommand{\section}{\@startsection{section}{1}{0em}%
        {-3.5ex \@plus -1ex \@minus -.2ex}%
        {2.3ex \@plus.2ex}%
        {\normalfont\large\bfseries}}
\renewcommand{\subsection}{\@startsection{subsection}{2}{0em}%
        {-3.25ex\@plus -1ex \@minus -.2ex}%
        {1.5ex \@plus .2ex}%
        {\normalfont\bfseries}}
\renewcommand{\subsubsection}%
        {\@startsection{subsubsection}{3}{0em}%
        {-3.25ex\@plus -1ex \@minus -.2ex}%
        {1.5ex \@plus .2ex}%
        {\normalfont\itshape}}
\makeatother

\newcommand{\be}[1]{\begin{equation} #1 \end{equation}}

\newcommand{\bin}[1]{b_{#1}^{N}}

\newcommand{\MGUTz}{M^0_\text{GUT}}
\newcommand{\GSM}{G_\text{SM}}
\newcommand{\GPS}{G_\text{PS}}
\newcommand{\GLR}{G_{LR}}

\newcommand{\SISSA}{SISSA/ISAS and INFN, I--34013 Trieste, Italy}

\newcommand{\preprintnumber}{%
SISSA--74/2008/EP}
 
\newcommand{\titletext}{Gauge coupling unification, the GUT scale, and magic fields} 
\newcommand{\authortext}{\large L.~Calibbi, L.~Ferretti, A.~Romanino and R.~Ziegler
\medskip\\\em\normalsize 
\SISSA}
\newcommand{\abstracttext}{We consider field sets that do not form complete SU(5) multiplets, but  exactly preserve the one-loop MSSM prediction for $\alpha_3(M_Z)$ independently of the value of their mass. Such fields can raise the unification scale in different ways, through a delayed  convergence of the gauge couplings, a fake unified running below the GUT scale, or a postponed unification after a hoax crossing at a lower scale. The $\alpha_3(M_Z)$ prediction is independent of the mass of the new fields, while the GUT scale often is not, which allows to vary the GUT scale. Such ``magic'' fields represent a useful tool in GUT model building. For example, they can be used to fix gauge coupling unification in certain two step breakings of the unified group, to suppress large KK thresholds in models with extra dimensions, or they can be interpreted as messengers of supersymmetry breaking in GMSB models.}

\title{
\normalsize
\hspace*{\fill}
\begin{tabular}[t]{l}\preprintnumber\end{tabular}
\vspace{3\baselineskip}\\\Large\bfseries\titletext\bigskip}
\author{\begin{minipage}[t]{0.8\textwidth}
\normalsize\centering\authortext
\end{minipage}}
\date{}

\begin{document}

\bigskip
\maketitle
\begin{abstract}\normalsize\noindent
\abstracttext
\end{abstract}\normalsize\vspace{\baselineskip}

\section{Introduction}

Gauge coupling unification can be considered as one of the most striking successes of the Minimal Supersymmetric Standard Model (MSSM). Together with the understanding of the pattern of the SM fermion gauge quantum numbers within Grand Unified Theories (GUTs), it represents one of the most convincing and quantitatively precise hints of physics beyond the SM model. The possibility to account for gauge coupling unification translates into a prediction of the strong coupling $\alpha_3 = g^2_3/(4\pi)$ in good agreement with the measured value, within the uncertainties associated to low energy and (unknown) high energy thresholds. The scale $\MGUT$ at which the couplings unify is also predicted to be $\MGUTz\approx 2\cdot 10^{16}\GeV$.

The MSSM prediction assumes that no additional fields appear in the spectrum before the unification scale: the so-called ``desert''. The study of the impact on gauge coupling unification of new fields with a mass between $\MGUT$ and the electroweak scale has a long history~\cite{Martin:1995wb,string,splitandref} and is based on at least two motivations. First, the appearance of new fields at a scale lower than $\MGUT$ is predicted by many theories beyond the SM. Since the unified gauge group is broken below $\MGUT$, such new fields will in general spoil gauge coupling unification. With enough parameters around, the latter can often be fixed, but only at the price of loosing what in the MSSM is an insightful prediction. As an example, neutrino masses are associated to lepton number violation, which can be due to right-handed neutrinos living one to five orders of magnitude below $\MGUTz$. When right-handed neutrino masses are associated with the breaking of the SU(2)$_R$ subgroup of SO(10), unification is affected, but can be fixed with an appropriate choice of the scale of right-handed neutrinos \cite{senjanovic}. 

The second important motivation for studying the effect of new fields on unification is raising the GUT scale $\MGUT$. The value of $\MGUT$ is crucial for proton decay. Within the $R$-parity conserving MSSM, proton decay is induced by dimension five and six operators. While dimension six operators are typically harmless, the decay rate induced by dimension five operators is often close or even above the experimental limit, depending on the embedding into the unified theory. For example, the minimal embedding into SU(5) is already excluded by the present bound on proton lifetime ~\cite{Hayato:1999az,Murayama:2001ur} for $\MGUT = \MGUTz$ (see however \cite{perez}). Raising the GUT scale is also useful in many string theory models in which the GUT scale turns out to be one order of magnitude larger than $\MGUTz$ \cite{string}. 

It is well known that fields forming complete SU(5) multiplets do not affect the prediction of $\alpha_s$ at the one loop level, independently of the scale at which they are added. This is useful but it does not address the above motivations. In particular, their presence does not affect $\MGUTz$ (at the one-loop level). In this short note we discuss what we call ``magic'' sets of fields. These are sets of vectorlike SM chiral superfields that do not form full SU(5) multiplets but share their benefits regarding gauge coupling unification: i) they exactly preserve the 1-loop MSSM prediction for $\alpha_3$ and ii) they do it independently of the value of their (common) mass. In particular, they maintain the predictivity of the MSSM, in the sense that their mass does not represent an additional parameter that can be tuned in order to fix $\alpha_3$ (at one loop). At the same time, magic sets do not form full SU(5) multiplets and therefore typically do have an impact on $\MGUT$, which is larger the further away they are from $\MGUTz$. In particular, there are several magic field sets that raise the GUT scale.

The paper is organized as follows. In Section~\ref{secmag}, we define the magic fields and discuss their effect on the GUT scale. In Section~\ref{sec:origin} we show that such fields can indeed be obtained from a unified theory. In Section~\ref{sec:PS} we consider the case in which the unified group SO(10) is broken in two steps, so that the gauge group below the unification scale is not the SM one. In Section~\ref{sec:applications} we discuss a few applications of magic fields. In particular, we consider the possibility to suppress Kaluza-Klein threshold effects in the context of unified theories with extra dimensions and gauge mediated supersymmetry breaking models with magic messengers. In the Appendix, we give systematic lists of magic fields. 

\section{Magic fields}
\label{secmag}

We consider the MSSM with additional vectorlike matter superfields at a scale $Q_0 > M_Z$. Let us denote by $b_i$, $i=1,2,3$ the 1-loop beta function coefficients for the three SM gauge couplings. At scales $M_Z < \mu < Q_0$, the MSSM spectrum gives $(b_1,b_2,b_3) = (33/5, 1, -3)\equiv(b^0_1, b^0_2, b^0_3)$. At $\mu > Q_0$, the beta coefficients include the contribution $b^N_i$ of the new fields, $b_i = b^0_i + b^N_i$ and the 1-loop running gives
\be{\frac{1}{\alpha_i(\mu)}=\frac{1}{\alpha_i(M_Z)}-\frac{b_i^0}{2\pi}\log\left(\frac{\mu}{M_Z}\right)-\frac{\bin{i}}{2\pi}\log\left(\frac{\mu}{Q_0}\right).}
The MSSM 1-loop prediction for $\alpha_3$, 
\begin{equation}
\label{eq:alpha3}
\frac{1}{\alpha_3} = \frac{1}{\alpha_2} 
+ \frac{b^0_3 - b^0_2}{b^0_2 - b^0_1} \left(\frac{1}{\alpha_2} - \frac{1}{\alpha_1}\right)
\end{equation}
is exactly preserved independently of the scale $Q_0$ if \cite{Martin:1995wb} 
\be{ \label{magcond} \frac{b_3^N-b_2^N}{b_2^N-b_1^N}=\frac{b_3^0-b_2^0}{b_2^0-b_1^0}=\frac{5}{7}.}
In this case, the unification scale becomes
\be{\MGUT = \MGUTz {\left( \frac{Q_0}{\MGUTz} \right)}^r\label{scale},}
with \be{ \label{rdef} r = \frac{\bin{3}-\bin{2}}{b_3-b_2},}
and the unified gauge coupling is given by
\be{\frac{1}{\alpha_U}=\frac{1}{\alpha_U^0}-\frac{(1-r) b_i^N-r b_i^0}{2\pi}\log\left(\frac{\MGUTz}{Q_0}\right)\label{aeq},}
where $\alpha^0_U$ is the value in the MSSM, $\alpha^0_U \sim 1/24$.
Complete GUT multiplets give the same contribution to the three beta functions and thus trivially satisfy eq. (\ref{magcond}); they preserve gauge coupling unification and leave the GUT scale unchanged. We call ``magic'' all other vectorlike sets of fields that satisfy eq. (\ref{magcond}) and therefore preserve the 1-loop MSSM prediction for $\alpha_3$. They fall into two categories: those with $r=0$, which just mimic the effect of complete GUT multiplets and those with $r \neq 0$, which change the GUT scale according to eq. (\ref{scale}).

The parameter $r$ also determines the relative order of the three scales $Q_0$, $\MGUTz$ and $\MGUT$. There are five different possibilities: 
\begin{itemize}
\item 
$r=0$ $\Rightarrow$ $Q_0< \MGUTz=\MGUT$: {\bf standard unification}. \\
This corresponds to $b_3^N=b_2^N=b_1^N$. The GUT scale is unchanged. The new fields can form complete GUT multiplets, but not necessarily. 
\item 
$-\infty<r<0$ $\Rightarrow$ $Q_0< \MGUTz<\MGUT$: {\bf retarded unification}. \\
The new fields slow the convergence of the gauge couplings. The simplest example of magic fields leading to retarded unification is $\left(Q + \bar{Q}\right) + G$ \footnote{Here and below we denote the new fields according to their quantum numbers as in the Appendix in Table~\ref{tab:notations}}, which gives $(b_3^N,b_2^N,b_1^N) = (5,3,1/5)$ and $r=-1$. The running of the gauge couplings is shown in Fig~\ref{fig:retarded}. 
\begin{figure}[ht]
\begin{center}
\includegraphics[angle=-90, width=0.5\textwidth]{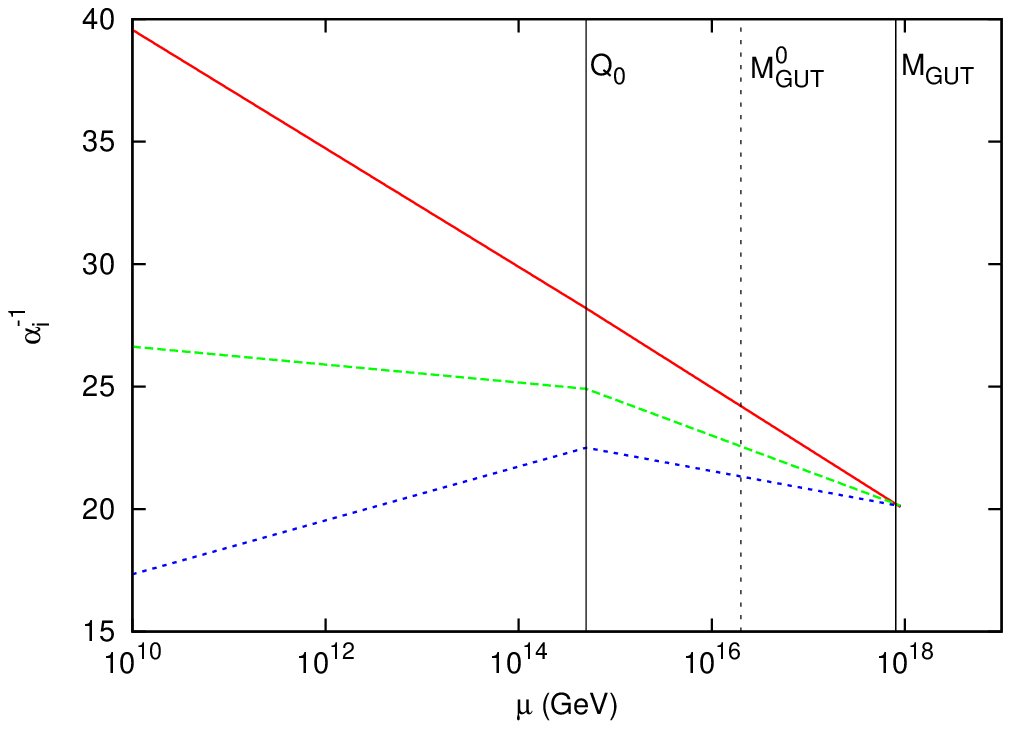}
\end{center}
\caption{Example of retarded unification. The fields $\left(Q + \bar{Q}\right) + G$ have been added at the scale $Q_0$.}
\label{fig:retarded}
\end{figure} 
\item 
$r=\pm\infty$ $\Rightarrow$ $Q_0=\MGUTz<\MGUT$: {\bf fake unification}. \\
This case corresponds to $b_3=b_2=b_1$. The unified group is broken at a scale $\MGUT\geq\MGUTz$, but the couplings run together between $Q_0 = \MGUTz$ and $\MGUT$, thus faking unification at the lower scale $\MGUTz$. Note that in this case $\MGUT$ is undetermined, while $Q_0$ is fixed.
A simple example can be obtained by adding the fields $(6,2)_{-1/6} + \text{c.c.}$\footnote{This representation is for example contained in the $\bf 210$ of SO(10).}, which gives $(b_3^N,b_2^N,b_1^N)=(10,6,2/5)$ (see Fig~\ref{fig:fake}). This possibility was previously considered in \cite{Dutta:2007ai}. 
\begin{figure}[ht]
\begin{center}
\includegraphics[angle=-90, width=0.5\textwidth]{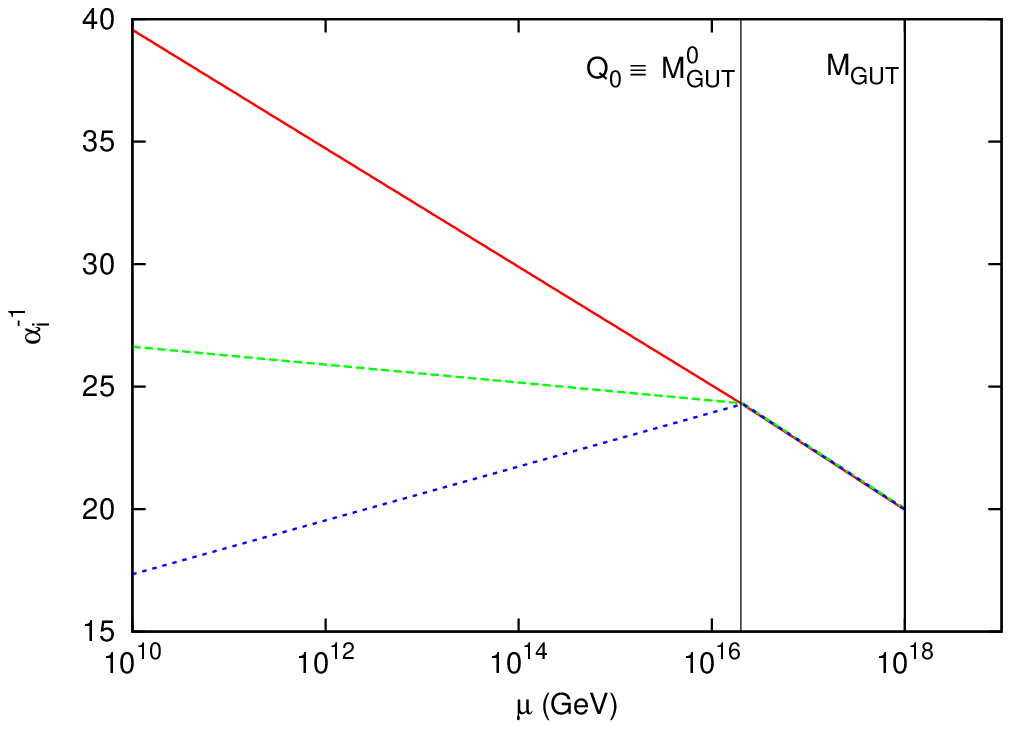}
\end{center}
\caption{Example of fake unification. The fields $(6,2)_{-1/6} + \text{c.c.}$ have been added at the scale $Q_0 = \MGUTz$.}
\label{fig:fake}
\end{figure}
\item 
$1<r<+\infty$ $\Rightarrow$ $\MGUTz<Q_0<\MGUT$ : {\bf hoax unification}. \\
In this scenario the magic set turns a convergent running into a divergent one and vice versa. Therefore such a field content cannot be added at a scale smaller than $\MGUTz$, or the gauge couplings would diverge above $Q_0$ and never meet. However unification is preserved if the magic fields are heavier than $\MGUTz$. Then the couplings, after an hoax crossing at $\MGUTz$, diverge between $\MGUTz$ and $Q_0$, start to converge above $Q_0$ and finally unify at $\MGUT$, the scale where the unified group is broken. 
For example,  the fields $W + 2 \times \left((8,2)_{1/2} + \text{c.c.} \right)$ \footnote{$\left((8,2)_{1/2} + \text{c.c.} \right)$ is contained both in the $\bf 120$ and $\bf 126$ of SO(10).} give $(b_3^N,b_2^N,b_1^N)=(24,18,48/5)$ and $r=3$ (see Fig~\ref{fig:hoax}).
\begin{figure}[ht]
\begin{center}
\includegraphics[angle=-90, width=0.5\textwidth]{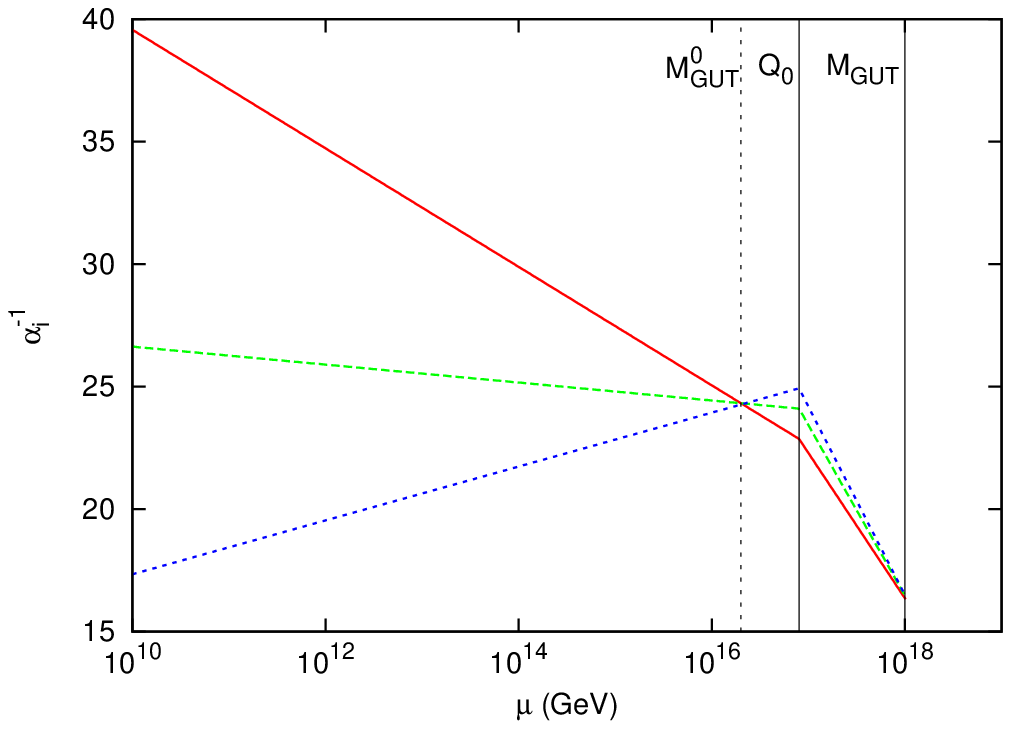}
\end{center}
\caption{Example of hoax unification.The fields $(1,3)_0 + 2 \times \left((8,2)_{1/2} + \text{c.c.} \right)$ have been added at the scale $Q_0 > \MGUTz$.}
\label{fig:hoax}
\end{figure}
\item $0<r<1$ $\Rightarrow$ $Q_0<\MGUT < \MGUTz$: {\bf anticipated unification}. \\
The magic content accelerates the convergence of the gauge couplings and the unification takes place before the usual GUT scale. This possibility can be useful in combination with other types of magic sets at different scales. 
\end{itemize}

Some comments are in order:
\begin{itemize}
 \item 
In the above considerations, the scale $Q_0$ is arbitrary, as long as $\MGUT \lesssim M_\text{Pl}$ and the unified gauge coupling is in the perturbative regime, $\alpha_U \lesssim 4 \pi$. 
\item
If we restrict our analysis to representations that can be obtained from the decomposition of SU(5) multiplets under $\GSM$, then both $b_3^N-b_2^N$ and $\frac{5}{2}(b_2^N-b_1^N)$ are integers. In this case the magic condition requires $b_3^N-b_2^N$ to be even and $b_2^N-b_1^N$ to be a multiple of $14/5$ \cite{Martin:1995wb}.
Therefore in the case of retarded unification the only possibility is $b_3^N-b_2^N=2$, which corresponds to $r=-1$. The expression for the GUT scale (\ref{scale}) becomes particularly simple: 
\begin{equation}
\frac{\MGUT}{\MGUTz} = \frac{\MGUTz}{Q_0} .
\end{equation} 
Therefore, in this scenario, $Q_0$ cannot be lower than $10^{13}-10^{14}$ GeV, in order to keep $\MGUT \lesssim M_\text{Pl}$. 
\item
An important property following from eq.~($\ref{magcond}$) is that combinations of magic sets at different scales do not spoil unification. In particular, merging two or more sets at the same scale gives again a magic set. Two simple rules are: adding two retarded solutions gives a fake solution, and adding a fake to a retarded solution or to another fake gives a hoax solution\footnote{Note that the classification based on $r$ can be rewritten in terms of the parameter $q=b_3^N-b^N_2$ used by \cite{Martin:1995wb}. Anticipated unification then corresponds to $q<0$, standard unification to $q=0$, retarded to $q=2$, fake to $q=4$, and hoax to $q>4$. The $q$ of a combination of magic fields sets is the sum of the individual $q$'s, from which the rules follow trivially.}.

\end{itemize}

\section{The origin of magic fields}
\label{sec:origin}

In this Section we show that magic field sets at a scale $Q_0 < \MGUT$ can indeed arise from the spontaneous breaking of a supersymmetric SO(10) GUT at the scale $\MGUT$. We will illustrate this in three examples for the case of retarded, fake, and hoax unification
\bigskip

\noindent
\textbf{1.} The simplest example of a magic field content leading to retarded unification is $ \left(Q + \bar{Q}\right) + G$, which can be obtained by splitting the components of a $16+\overline{16}+45$ of SO(10). As an example, such a splitting is provided by the following superpotential:
\begin{multline}
\label{eq:ret-sp}
W = 
16\ 45_H \overline{16}+16_H\ 16\ 10+\overline{16}_H\ \overline{16}\ 10+45_H\ 45\ 54 \\ +16_H\ 45\ {\overline{16}}^\prime+\overline{16}_H\ 45\ 16^\prime + M\ 10\ 10 + M\ 54\ 54 + M\ {\overline{16}}^\prime 16^\prime.
\end{multline}
Here and below, all dimensionless couplings are supposed to be $\ord{1}$ and $M\approx \MGUT$. The $45_H$ is assumed to get a vev of order $\MGUT$ along the $T_{3R}$ direction. Then, the above superpotential gives a mass of order $\MGUT$ to all fields except $Q$, $\bar{Q}$, $G$, which are assumed to get a mass at a lower scale $Q_0$. A two-loop analysis shows that the prediction for $\alpha_3(M_Z)$ does not significantly differ from the MSSM one. 
\bigskip

\noindent
\textbf{2.} An example of fields leading to fake unification is 
$$2\times (L+\bar{L})+2\times G+2\times W+2\times (E+\bar{E})+\left((8,2)_{1/2}+\text{c.c}\right),$$ 
which can be embedded into a $45+45+120$ of SO(10). 
This magic field set can be obtained from the following superpotential 
\be{W=45\ 45_H\ 45'+120\ 45_H\ 120'+M\ 120'\ 120',} 
if $45_H$ gets a vev of order $M \approx \MGUT$ along the $B-L$ direction.
Another example is $2\times (Q+\bar Q+G)$, which can be obtained by a generalization of the superpotential in \eq{ret-sp} 
\bigskip

\noindent
\textbf{3.} As an example for hoax unification, we consider the set $4 \times (L + \bar{L}) +3 \times \left((8,2)_{1/2}+\text{c.c} \right)$, which can be embedded into a $120+2 \times (126 + \overline{126})$. An example of superpotential is
\be{W=126\ 45_H\ \overline{126}+126'\ 45_H\ \overline{126}'+120\ 45_H\ 120'+M\ 120'\ 120',}
again with the vev of $45_H$ along the $B-L$ direction.

\section{Magic field content in a 2-step breaking of SO(10)}
\label{sec:PS}

The necessity of achieving gauge coupling unification in the presence of fields not forming full unified multiplets may arise in the context of a two-step breaking of SO(10): SO(10) is broken at the scale $\MGUT$ to the intermediate group $G_i$, which is then broken to the SM at a lower scale $M_i$. In fact, the presence of an intermediate gauge group at a lower scale $M_i < \MGUT$ often spoils gauge unification if no further fields are added. This is because the additional gauge bosons of $G_i/\GSM$ are not necessarily in full SU(5) multiplets, as in the case of $G_i = \GPS \equiv \text{SU(2)}_L \times \text{SU(2)}_R \times \text{SU(4)}_c$ and $G_i = \GLR \equiv \text{SU(2)}_L \times \text{SU(2)}_R \times \text{SU(3)}_c \times \text{U(1)}_{B-L}$. 

Let us consider a set of fields at the scale $Q_0$, with $M_i < Q_0 < \MGUT$, which consists of multiplets of the gauge group $G_i$. The condition~(\ref{magcond}) for preserving unification changes, since it has now to take into account the additional vector superfields and has to be expressed in terms of the beta coefficients of the gauge couplings of the group $G_i$. 

Let us first consider the case $G_i = \GPS$. We denote a PS multiplet by $(a,b,c)$, where $a,b,c$ are the quantum numbers under SU(4)$_c$, SU(2)$_L$, SU(2)$_R$ respectively. The three PS gauge couplings $g_4$, $g_L$, $g_R$ are matched to the SM ones at the PS breaking scale $M_\text{PS}$ as follows:
\begin{equation}
\label{eq:gaugematching}
\frac{1}{\alpha_4} = \frac{1}{\alpha_3}
\qquad
\frac{1}{\alpha_L} = \frac{1}{\alpha_2}
\qquad
\frac{1}{\alpha_R} = \frac{5}{3} \frac{1}{\alpha_1} -\frac{2}{3} \frac{1}{\alpha_3} .
\end{equation}
In terms of their beta function coefficients $b_4$,$b_L$, $b_R$, the condition~(\ref{magcond}) becomes
\be{\label{eq:magic-ps}\frac{b_4-b_L}{b_L-b_R}=\frac{1}{3} .}
The contribution of MSSM fields and PS gauge bosons is $(b_4^0,b_L^0,b_R^0)=(-6,1,1)$. Thus, the Pati-Salam couplings do not unify if no extra matter is added, as condition ~(\ref{eq:magic-ps}) is not satisfied. It is however possible to restore unification by adding a single $(6,1,3)$ field at the PS breaking scale, which exactly cancels the contribution of the massive PS gauge bosons to the beta function coefficients. Note that extra matter is also needed in order to break Pati-Salam to the SM. 

If the field content below the PS scale is the MSSM one, the classification given in section \ref{secmag} can be maintained in these models simply by replacing $r$ in eq.~(\ref{rdef}) with 
\be{r=\frac{b_4^N-3-b_L^N}{b_4-b_L}\label{r2} .}
The formula (\ref{scale}) for the GUT scale is then still valid. A more general expression for the new unification scale valid whatever is the (magic) field content below $M_i$ is
\begin{equation}
\ln\frac{\MGUT}{\MGUTz}=\left(\frac{b_3-b_2}{b_4-b_L}-1\right)\ln\frac{\MGUTz}{M_\text{PS}},
\end{equation} 
where $b_2,b_3$ are the MSSM beta coefficients just below the PS scale.

There are simple examples of magic fields that get their mass from PS-breaking vevs (the fields getting vev are also assumed to be part of a unified or magic multiplet). One example leading to retarded unification is $(4,1,2)+(\bar{4},1,2)+(1,2,2)+(1,1,3)+(10,2,2)+(\overline{10},2,2)$, with masses obtained from a $(15,1,1)$ vev along the $B-L$ direction. An example for fake unification is $(6,1,1)+(10,1,1)+(\overline{10},1,1)$, with masses obtained from a $(1,1,3)$ vev proportional to $T_{3R}$. 

An example for fake unification, which also accounts for the Pati-Salam breaking, can be constructed by using the fields
\begin{equation*}
A(6,1,1)+W_4(15,1,1)+ \left[S(10,1,1)+S_L(10,3,1)+S_R(10,1,3)
 +F(4,2,1)+F^c(\bar{4},1,2)+ \text{c.c.}\right], 
\end{equation*}
and the superpotential
\begin{multline}
\label{eq:ps-sp}
W=\bar{F}^c W_4 F^c+ S_R F^c F^c +\bar{S}_R \bar{F}^c \bar{F}^c + M_F \bar{F}^c F^c + M_{S}\bar{S}_R S_R + \frac{M_W}{2} W_4 W_4 \\  + W_4 A A + \bar{S} W_4 S+ \bar{S}_L W_4 S_L+ \bar{F} W_4 F ,
\end{multline}
where all the dimensionless couplings are of order 1 and the masses are of order $M_\text{PS} = \MGUTz$. The interactions in the first line of ~(\ref{eq:ps-sp}) break PS to SM giving non-zero vevs to $S$, $F^c$, $W_4$ and their conjugates, while those in the second line give mass to all the other fields.
Note that in this case the Pati-Salam breaking scale corresponds to the gauge coupling unification scale, while SO(10) is broken at an higher scale $\MGUT$ that is undetermined.

As for the case $G_i = \GLR$, the magic condition can be written in terms of the beta coefficients $b_L$, $b_R$, $b_3$, $b_{B-L}$ as
\be{ \label{magcondlr} \frac{b_3-b_{2L}}{b_{2L}-\frac{3}{5}b_{2R}-\frac{16}{15}b_{B-L}}=\frac{5}{7}.}
The contribution of the MSSM and the additional $\GLR$ gauge bosons to the beta coefficients is $(b_L, b_R, b_3, b_{B-L}) = (1,1,-3,16)$ and the expression for $r$  is the same as in the MSSM (see eq.~(\ref{rdef})), with $b_2 = b_L$. Some examples which are related to the discussion in this Section can be found in \cite{Bando:1994hg}.

\section{Applications}
\label{sec:applications}

\subsection{The magic tower}

An interesting application arises in unified theories with extra dimensions compactified on an orbifold. 
The advantages of such orbifold GUTs have been widely discussed in the literature and include easy breaking of the unified group by orbifold boundary conditions, a straightforward solution of the 2-3 splitting problem, and the suppression of dangerous baryon number violating dimension-five operators~\cite{SU5OFGUTS}. In these theories, fields living in the bulk of the extra dimension correspond to ``Kaluza-Klein'' (KK) towers of fields in the effective four-dimensional theory, whose masses are multiples of the compactification scale. Because of the very mechanism of GUT breaking by orbifolding, the KK fields with a given mass do not form full multiplets of the unified group. As a consequence, the KK towers associated to the bulk fields introduce new thresholds affecting the prediction of $\alpha_3$. While such thresholds are often used to improve the agreement with data (if they are not too large), it is interesting to note that it is possible to get rid of such effects if the fields corresponding to a given KK mass form magic sets. 

As an example, let us consider a 5D supersymmetric SO(10) model on $S^1/(Z_2\times Z_2')$ with a Pati-Salam brane and a SO(10) brane (see~\cite{SO10OFGUTS} for a description of such models). The vector fields $(V,\Sigma)$ live in the bulk together with a chiral hypermultiplet $(\Phi_1,\Phi_2)$ in the adjoint of SO(10), while the SM matter, the Higgses and other fields live on the branes. The bulk fields can be classified in terms of their two orbifold parities $(P_1,P_2) = (\pm 1, \pm 1)$. The orbifold boundary conditions are chosen such that the SO(10) adjoints $V$, $\Sigma$, $\Phi_1$, $\Phi_2$ split into their PS adjoint components and the orthogonal component, with orbifold parities defined as follows
\begin{center}
\begin{tabular}{c|c|c}
$(V,\Sigma)$ &$(\Phi_1,\Phi_2)$ &\\
\hline
$V_{++},\Sigma_{--}$ &$\Phi_{1++},\Phi_{2--}$ & PS adjoints\\
$V_{+-},\Sigma_{-+}$ &$\Phi_{1+-},\Phi_{2-+}$ & SO(10)/PS adjoints 
\end{tabular} .
\end{center}
The massless zero-modes are given by the gauge fields $V^{(0)}_{++}$ and an adjoint field $\Phi^{(0)}_{1++}$. 
The odd KK states contain fields of the SO(10)/PS adjoint representation, while the even KK states contain those of the PS adjoint.

Clearly, neither the even nor the odd states correspond to full SO(10) (or SU(5)) multiplets. Still, both of them could form magic sets, in which case the threshold effects associated to the KK tower of fields would vanish at one loop. This is indeed the case in the example we are considering, The easiest way to see it is to observe that the $(V,\Sigma)$ and $(\Phi_1,\Phi_2)$ multiplets together form an $\mathcal{N}=4$ SUSY hypermultiplet, which gives no contribution to the beta functions (the contribution of three chiral multiplets $\Sigma,\Phi_1,\Phi_2$ cancels exactly the one of the gauge fields $V$). Therefore both the even and the odd levels of the KK towers do not spoil unification.

Since we have not observed it, the zero-mode $\Phi_{1++}$ cannot be too light. It should have a mass at some intermediate scale $M_{\Phi}$, which can be identified with the PS breaking scale. In order to maintain unification it is sufficient to add some fields of mass $M_{\Phi}$ on the PS brane which form a magic field content together with $\Phi_{1++}$: an example is (4,1,2)+(6,1,1)+(1,1,3).

\subsection{Multi-scale models}

We briefly discuss an example of a model with multiple intermediate scales and a magic content of fields at all scales, which puts together many of the ideas discussed above and illustrates the property that different sets of magic fields can be added at different scales without spoiling unification. This is a modified version of the flavor model based on the Pati-Salam gauge group in~\cite{Ferretti:2006df}, in which the magic fields correspond to flavour messengers. The quantum numbers of the chiral supermultiplets of the model are given in Table~\ref{tab:PS}. There, $f_i = (l_i,q_i), f_i^c = (n^c_i,e^c_i,u_i^c,d_i^c), h = (h_u,h_d)$ contain the MSSM fields and $F+\bar{F}$, $F_c+\bar{F}_c$ is an heavy vector-like copy of one SM generation. We call $A_\Phi,T_\Phi,\bar{T}_\Phi,G_\Phi$ the SM components of the SU(4) adjoint field $\Phi$.
\begin{table}[t]
\begin{equation*}
\begin{array}{|c|cccc|cccc|cccc|ccc|}
\hline
& f_i & f_i^c & h & \phi & F & \bar F & F^c & {\bar F}^c & F'_c & {\bar F}'_c & X_c & \Phi & H & \phi_L & \phi_R\\ \hline
\text{SU(2)}_L &  2 & 1 & 2 & 1 & 2& 2 &1 & 1 &1 &1 &1 &1 & 2 & 3 & 1\\
\text{SU(2)}_R &  1& 2& 2& 1& 1& 1& 2& 2& 2& 2& 3& 1 &2 &1 &3 \\
\text{SU(4)}_c &  4& \bar{4} &1 & 15 &4& \bar{4} & \bar{4} &4& \bar{4} &4& 1&15 &1 &1 &1\\
\hline
\end{array}
\end{equation*}
\caption{Quantum numbers of the fields of the Pati-Salam model.}
\label{tab:PS}
\end{table}
The model is characterized by two scales, $M_L$, $M_R$, with $M_L \ll M_R$. The Pati-Salam gauge symmetry is broken at $M_R$ to the SM one. The matter content at different mass scales $\mu$ is
\begin{itemize}
\item $\mu < M_L$: the usual MSSM field content;
\item $M_L< \mu <M_R$: the MSSM fields, the left-handed fields $F+\bar{F}$, $\phi$ and the color octet $G_\Phi$. A magic field set is obtained by adding the fields $H,\phi_L,\phi_R$ in the last block of the table;
\item $\mu > M_R$: all the fields in the table. 
\end{itemize}

The field content is magic at all scales. The one at $M_L$ corresponds to a retarded solution and therefore the unification scale is raised according to
\begin{equation}
\MGUT=\MGUTz \frac{\MGUTz}{M_L},
\end{equation} 
while the field content at $M_R$ does not modify the GUT scale.

This model can be embedded in a 5D supersymmetric GUT theory on a $S^1/(Z_2\times Z_2')$ orbifold. The fields $f_i,f^c_i$ live on the SO(10) brane, $F_c',\bar{F}_c',X_c$ on the PS brane and all the other fields in the bulk. In this setup all the KK levels turn out to be magic, giving rise to a nontrivial example of the magic KK towers discussed in the previous section. A very similar model was considered in \cite{CFRZflavour}.

\subsection{Gauge mediation}

In gauge mediated supersymmetry breaking (GMSB), the messenger sector is usually assumed to be made of full SU(5) multiplets in order not to spoil gauge coupling unification. In the light of the above discussion, it is natural to consider also the case of a messenger sector consisting of magic field sets. 
Gauge mediation with incomplete GUT multiplets was studied in \cite{Martin:1996zb}, and many of the conclusions apply also to this case. However, the requirement of gauge coupling unification gives additional constraints on the sparticle spectrum.

We assume the usual superpotential
\be{W=S\bar{\Psi}_i\Psi_i+M\bar{\Psi}_i\Psi_i,}
where $\Psi_i,\bar{\Psi}_i$ form a magic set of fields and $S$ is the spurion with $\langle F_S \rangle \neq 0$.
The gaugino masses at the scale $\mu$ are given by
\be{M_a(\mu)=\frac{\alpha_a(\mu)}{4\pi}b_a^N\frac{F_S}{M},}
while the scalar masses are
\be{\tilde{m}_i^2(\mu)=\sum_a 2 \left(\frac{\alpha_a(\mu)}{4\pi}\right)^2 C_a^i b_a^N \left[\frac{\alpha_a^2(Q_0)}{\alpha_a^2(\mu)}-\frac{b_a^N}{b^0_a}\left(1-\frac{\alpha_a^2(Q_0)}{\alpha_a^2(\mu)}\right)\right] \left|\frac{F_S}{M}\right|^2 ,}
where $C_a^i$ is the quadratic Casimir, $a$ is the index of the gauge group, $i$ runs over the matter fields, and $b_a^N$ is the contribution from the messengers to the beta function coefficients.
On the basis of the above expression, the sum rules on sfermion masses that hold in gauge mediation models \cite{Martin:1996zb,Cohen:2006qc}
are still valid. Interestingly, we obtain a sum rule for gaugino masses valid at all scales, which reads
\be{7 \frac{M_3}{\alpha_3}-12  \frac{M_2}{\alpha_2}+5 \frac{M_1}{\alpha_1}=0 .}
Typically, gaugino and scalar mass hierarchies turn out to be more pronounced than in the usual scenario. For instance, if the messenger sector is given by $Q+\bar{Q}+G$, the ratio between gaugino masses is very peculiar, $M_1:M_2:M_3=1:30:200$, and also the scalar masses turn out to be quite split: $m_{\tilde e^c}/m_{\tilde q}\sim 1/20$. For a less peculiar scenario such as $(Q+\bar{Q}) + G + (U^c+\bar{U}^c) + (D^c + \bar{D}^c) + W$, we get $M_1:M_2:M_3=1:5:20$ and $m_{\tilde e^c}/m_{\tilde q}\sim 1/15$. 
 An example of a typical SUSY spectrum for the two retarded solutions above, with the selectron mass taken close to the present experimental limit is 
\begin{equation*}
\begin{array}{|c|ccc|cc|}
\hline
 & M_1 & M_2 & M_3 & m_{\tilde{e}^c} & m_{\tilde{q}} \\
\hline
Q\bar{Q}+G & 25\ \GeV & 750\ \GeV & 5\ \TeV & 100\ \GeV & 2\ \TeV \\
Q\bar{Q}+G+U^c\bar{U}^c+D^c\bar{D}^c+W & 75\ \GeV & 400\ \GeV & 1.5\ \TeV & 100\ \GeV & 1.5\ \TeV \\ 
\hline
\end{array}
\end{equation*}
For messenger sectors with $b_1^N=b_2^N=b_3^N$, the spectrum of soft masses is the same as in the usual minimal gauge mediation scenario with an effective number of $5+\bar{5}$ messengers given by $b_i^N$.

\section{Summary}

In this note we systematically analyzed ``magic'' fields, sets of SM chiral superfields that do not form complete SU(5) multiplets, but exactly preserve the 1-loop MSSM prediction for $\alpha_3(M_Z)$ independently of the value of their mass. Unlike full SU(5) multiplets, such magic field sets may have an impact on the GUT scale.  In particular, we have shown that $\MGUT$ can be increased in three ways, through a delayed convergence of the gauge couplings, a fake unified running of the gauge couplings below the GUT scale, or a late unification after an hoax crossing of the gauge couplings at a lower scale. We have also shown several examples of dynamics giving rise to magic field contents below the unification scale. 

Increasing the unification scale is useful to suppress the proton decay rate and to make $\MGUT$ closer to the string scale. As the MSSM $\alpha_3(M_Z)$ prediction is not changed (at one loop) whatever is the scale $Q_0$ at which the new fields are added, the effect on the GUT scale can be tuned by varying $Q_0$, while maintaining predictivity on $\alpha_3(M_Z)$. 

Magic fields can have several applications. They can fix gauge coupling unification in two step breakings of the unified group by compensating the effect of the additional gauge bosons at the intermediate scale on the prediction for $\alpha_3(M_Z)$. They can be used to suppress too large thresholds from KK towers in models in which unification is achieved in extra dimensions. Or they can be interpreted as messengers of supersymmetry breaking in GMSB models. In summary, they represent a useful tool in GUT model building. 

\section*{Acknowledgments}
This work was supported by the EU FP6 Marie Curie Research \& Training Network ``UniverseNet'' (MRTN-CT-2006-035863)

\section*{Appendix: some magic field contents}

In this Appendix we show the results of a systematic analysis of magic field contents. Note that merging two or more magic sets still gives a magic set of fields. In particular, adding a magic content with $r=0$ does not modify the type of unification; adding two retarded solutions gives a fake solution, and adding a fake to a retarded solution or to another fake gives a hoax solution.
Table~\ref{tab:1} contains the simplest irreducible magic sets that can be built from SM representations belonging to SO(10) representations with dimension up to \textbf{210}. The notation for these representations is explained in Table~\ref{tab:notations}. We have not included field sets that form complete SU(5) multiplets. Table~\ref{tab:2} shows the simplest irreducible magic sets which provide retarded unification. Table~\ref{tab:3} shows the simplest irreducible magic contents for the Pati-Salam case. Again we write only fields belonging to representations of SO(10) up to \textbf{210}.
\begin{table}
\begin{center}
\begin{tabular}{|c|ccc ccc ccc|}
\hline
 & $Q$ & $U^c$ & $D^c$ & $L$ & $E^c$ & $W$ & $G$ & $V$ & $(n,m)_y$  \\
\hline\hline 
SU(3)$_c$ & 3 & $\bar{3}$ &  $\bar{3}$ & 1 & 1 & 1 & 8 & 3 & $n$ \\
SU(2)$_L$ & 2 & 1 & 1 & 2 & 1 & 3 & 1 & 2 & $m$ \\
Y & 1/6 & -2/3 & 1/3 & -1/2 & 1 & 0 & 0 & -5/6 & $y$ \\
\hline\hline
\end{tabular}
\end{center}
\caption{SM quantum numbers associated to a given notation for a SM field.}
\label{tab:notations}
\end{table}
\begin{table}
\begin{center}
\begin{tabular}{|c|c c c|c|c|}
\hline
 Field content & $b_1^N$ & $b_2^N$ & $b_3^N$ & $r$ & type \\
\hline\hline
 $(6,2)_{-1/6} + {\rm c.c.} $ & 2/5 & 6 & 10 & $\infty$ & fake \\ 
\hline
 $ \left(Q + \bar{Q}\right) + G$ & 1/5 & 3 & 5 & -1 & retarded \\
\hline
 $ \left(U^c + \bar{U}^c\right) + \left(D^c + \bar{D}^c\right) + W$ & 2 & 2 & 2 & 0 & usual \\
 $ \left(D^c + \bar{D}^c\right) + G + \left((1,3)_{1} + {\rm c.c.} \right) $ & 4 & 4 & 4 & 0 & usual \\
 $  \left(L + \bar{L}\right)  + \left((6,1)_{1/3}+ (1,3)_{1} + {\rm c.c.}\right) $ & 5 & 5 & 5 & 0 & usual \\
 $ \left(Q + \bar{Q}\right) + \left(D^c + \bar{D}^c\right) +  \left((8,2)_{1/2} + {\rm c.c.}\right)$ & 27/5 & 11 & 15 & $\infty$ & fake \\
 $ W + 2 \left((8,2)_{1/2} + {\rm c.c.}\right)$ & 48/5 & 18 & 24 & 3 & hoax \\
 $ W + \left((6,2)_{-1/6} + {\rm c.c.}\right) + \left((1,1)_{2} + {\rm c.c.}\right) $ & 26/5 & 8 & 10 & -1 & retarded \\
 $  \left( (3,3)_{2/3} + (6,2)_{-1/6} + (6,1)_{4/3} + {\rm c.c.}\right)$ & 18 & 18 & 18 & 0 & usual \\
 $ 2 W +  \left((6,2)_{5/6} + {\rm c.c.} \right)$ & 10 & 10 & 10 & 0 & usual \\
 $ \left((3,3)_{2/3} + (6,2)_{5/6} + (6,1)_{-2/3} + {\rm c.c.} \right)$ & 18 & 18 & 18 & 0 & usual \\
 $ \left((8,1)_{1} + (\bar 3,1)_{4/3} + {\rm c.c.} \right) +  (8,3)_{0}$ & 16 & 16 & 16 & 0 & usual \\
 $ \left((8,1)_{1} + (6,1)_{1/3} + {\rm c.c.} \right) +  (8,3)_{0}$ & 52/5 & 16 & 20 & $\infty$ & fake \\ 
\hline
\hline
\end{tabular}
\end{center}
\caption{Simplest irreducible magic sets that can be built from SM representations belonging to SO(10) representations with dimension up to \textbf{210} and do not correspond to full SU(5) multiplets or anticipated unification.}
\label{tab:1}
\end{table}
\begin{table}
\begin{center}
\begin{tabular}{|c|c c c|c|}
\hline
 Field content & $b_1^N$ & $b_2^N$ & $b_3^N$ & $r$  \\
\hline\hline
 $ \left(Q + \bar{Q}\right) + G$ & 1/5 & 3 & 5 & -1  \\
\hline
$ \left(E^c + \bar{E}^c\right)+ 2 W + 2 G$ & 6/5 & 4 & 6 & -1  \\
$ 2 \left(L + \bar{L}\right)+  W + 2 G$ & 6/5 & 4 & 6 & -1  \\
$ \left(Q + \bar{Q}\right) + \left(U^c + \bar{U}^c\right)+ \left(D^c + \bar{D}^c\right)+W + G$ & 11/5 & 5 & 7 & -1  \\
\hline
$ 3 \left(D^c + \bar{D}^c\right) +2 W +  G $ & 6/5 & 4 & 6 & -1  \\
$ \left(U^c + \bar{U}^c\right) + \left(L + \bar{L}\right) + 2 W + 2 G $ & 11/5 & 5 & 7 & -1  \\
$ \left(Q + \bar{Q}\right) + 2\left(D^c + \bar{D}^c\right) + \left(E^c + \bar{E}^c\right) +  W +  G $ & 11/5 & 5 & 7 & -1  \\
$ 2\left(Q + \bar{Q}\right) + \left(D^c + \bar{D}^c\right) + 2\left(E^c + \bar{E}^c\right) +  G $ & 16/5 & 6 & 8 & -1  \\
$ 2\left(Q + \bar{Q}\right) + \left(U^c + \bar{U}^c\right) + 3\left(D^c + \bar{D}^c\right) $ & 16/5 & 6 & 8 & -1  \\
$ 2\left(Q + \bar{Q}\right) + 2\left(U^c + \bar{U}^c\right) + \left(L + \bar{L}\right)+G $ & 21/5 & 7 & 9 & -1  \\
$ 2\left(Q + \bar{Q}\right) + 2\left(D^c + \bar{D}^c\right) +G+ \left(V + \bar{V}\right) $ & 31/5 & 9 & 11 & -1  \\
\hline\hline
\end{tabular}
\end{center}
\caption{Simplest irreducible magic sets which provide retarded unification. We show only fields belonging to representations of SO(10) up to \textbf{45}.}
\label{tab:2}
\end{table}
\begin{table}
\begin{center}
\begin{tabular}{|c|c c c|c|}
\hline
 Field content & $b_4^N$ & $b_L^N$ & $b_R^N$ &  $r$  \\
\hline\hline
$(6,1,3)$ & 3 & 0 & 12 & 0 \\
\hline
$(1,2,2) + \left((20',1,1)+{\rm c.c.}\right)$ & 8 & 1 & 1 & $\infty$ \\
$(6,1,1) + \left((10,1,1)+{\rm c.c.}\right)$ & 7 & 0 & 0 & $\infty$ \\
$\left((10,1,1)+{\rm c.c.}\right) + (15,2,2)$ & 22 & 15 & 15 & $\infty$ \\
\hline
$(1,2,2) + 2 (15,1,1)$ & 8 & 1 & 1 & $\infty$ \\
$(6,1,1) + (6,2,2) + \left((20',1,1)+{\rm c.c.}\right)$ & 13 & 6 & 6 & $\infty$ \\
$(6,1,1) + (6,1,3) + (1,2,2)$ & 4 & 1 & 13 & 0 \\
$\left((4,1,2)+(4,2,1)+{\rm c.c.}\right) + (6,1,3)$ & 7 & 4 & 16 & 0 \\
$(1,3,3) + \left((10,1,1)+{\rm c.c.}\right) + (6,1,3)$ & 9 & 6 & 18 & 0 \\
$(6,2,2) + \left((20',1,1)+{\rm c.c.}\right) + (15,2,2)$ & 28 & 21 & 21 & $\infty$ \\
$(1,2,2) + (6,1,3) + (15,2,2)$ & 19 & 16 & 28 & 0 \\
$(1,1,3) + (6,1,3) + \left((20,2,1)+{\rm c.c.}\right)$ & 29 & 20 & 14 & -3 \\
$(6,1,3) + \left((4,2,3) + (20,2,1)+{\rm c.c.}\right)$ & 35 & 32 & 44 & 0 \\
$(6,1,3) + \left((4,3,2) + (20,1,2)+{\rm c.c.}\right)$ & 35 & 32 & 44 & 0 \\
$(6,2,2) + (6,3,1) + (15,1,3)$ & 19 & 18 & 36 & 1/3 \\
$(1,2,2) + (15,1,1) + \left((10,2,2)+{\rm c.c.}\right)$ & 28 & 21 & 21 & $\infty$ \\
$(1,2,2) + 2 \left((10,2,2)+{\rm c.c.}\right)$ & 48 & 41 & 41 & $\infty$ \\
\hline\hline
\end{tabular}
\end{center}
\caption{Simplest irreducible magic contents for the Pati-Salam case that can be built from PS representations belonging to SO(10) representations with dimension up to \textbf{210} and do not correspond to full SU(5) multiplets or anticipated unification. We denote the fields as $(a,b,c)$, where $a$, $b$, $c$ are representations of SU(4)$_c$, SU(2)$_L$, SU(2)$_R$ respectively.}
\label{tab:3}
\end{table}

\end{document}